\documentclass{sig-alternate-2013}

\usepackage{hypothesis}
\usepackage{booktabs}
\usepackage{graphicx}   

\usepackage{url}
\usepackage[font=bf,skip=\baselineskip]{caption}
\DeclareCaptionType{copyrightbox}
\usepackage{subcaption}

\newcommand{\eg}{e.g.~\ignorespaces}
\newcommand{\ie}{i.e.~\ignorespaces}
\newcommand{\thing}{paper}

\urldef{\wikisurvey2011}\url{https://meta.wikimedia.org/w/index.php?title=Editor_Survey_2011/Location_%26_Language&oldid=8409990} 

\clubpenalty=10000
\widowpenalty = 10000
\interfootnotelinepenalty=10000

\begin{document}
%

\newfont{\mycrnotice}{ptmr8t at 7pt}
\newfont{\myconfname}{ptmri8t at 7pt}
\let\crnotice\mycrnotice%
\let\confname\myconfname%

\toappear{\copyright~Scott A. Hale 2014. This is the author's version of the work. It is posted here for your personal use. Not for redistribution. The definitive version was published in WebSci '14, http://dx.doi.org/10.1145/2615569.2615684.}

\clubpenalty=10000
\widowpenalty = 10000


\title{Multilinguals and Wikipedia Editing
}

\numberofauthors{1} 
%
\author{
\alignauthor
Scott A.~Hale\\
       \affaddr{Oxford Internet Institute, University of Oxford}\\
       \affaddr{1 St Giles, Oxford, UK OX1 3JS}\\
       \email{scott.hale@oii.ox.ac.uk}
}
\date{23 February 2014}

\maketitle
\begin{abstract}
This article analyzes one month of edits to Wikipedia in order to examine the role of users editing multiple language editions (referred to as multilingual users). Such multilingual users may serve an important function in diffusing information across different language editions of the encyclopedia, and prior work has suggested this could reduce the level of self-focus bias in each edition. This study finds multilingual users are much more active than their single-edition (monolingual) counterparts. They are found in all language editions, but smaller-sized editions with fewer users have a higher percentage of multilingual users than larger-sized editions. About a quarter of multilingual users always edit the same articles in multiple languages, while just over 40\% of multilingual users edit different articles in different languages. When non-English users do edit a second language edition, that edition is most frequently English. Nonetheless, several regional and linguistic cross-editing patterns are also present.
\end{abstract}





\newpage
\section{Introduction}
Wikipedia, the free, peer-produced online encyclopedia, contains a large collection of human knowledge. The foundation behind Wikipedia has characterized the encyclopedia as trying to provide access to ``the sum of all human knowledge.''%
\footnote{\url{http://www.theatlantic.com/technology/archive/2011/05/is-wikipedia-a-world-cultural-repository/239274/}}
If any one language edition of Wikipedia were to achieve the goal of ``all human knowledge,'' then that language should contain (at a minimum) all the information found in other language editions of the encyclopedia. Studies comparing content across language editions, however, have found a ``surprisingly small amount of content overlap between languages of Wikipedia'' \cite[p.~295]{hecht2010}. No one edition contains all the information found in other language editions, and the largest language edition, English, contains only 51\% of the articles in the second-largest edition, German \cite[p.~295]{hecht2010}. Nonetheless, there clearly is some overlap in content between languages, and a greater sharing of information between the language editions would enable monolingual readers of the encyclopedia to access a larger variety of content.

This \thing{} examines one month of all edits to the top 46 language editions of Wikipedia. This comprises all editions with at least 100,000 articles at the time of data collection in July 2013.%
\footnote{\url{http://meta.wikimedia.org/wiki/List_of_Wikipedias}}
It identifies users who contribute to multiple language editions (these users are referred to as multilingual users in this \thing{}) and compares their contributions to that of users who edit only one language edition of the encyclopedia (monolingual users). It asks if multilingual editors play a unique bridging role diffusing information between different language editions.

\section{Related work}
Language is a large factor in the network structure of communication patterns on many platforms including  telephone communications \cite{barnett1995}, Twitter messaging \cite{eleta2012, hale-chi2014}, and blog linking \cite{hale-msc,herring2007}. Consistent with this previous work on other platforms, Hecht and Gergle~\citeyear{hecht2010} found there was low overlap in articles between different language editions of Wikipedia. In a separate study, they also found that most language editions exhibited a self-focus bias where articles about places, people, and events where the language of the edition was spoken were more prominent than those in other regions \cite{hecht2009}. While past work has not specifically looked at what percentage of users contribute to multiple language editions of the encyclopedia, studies of other platforms and the low overlap in content between different language editions of Wikipedia suggest that \hypothesis{h:wikihomophily}{most editors will edit only one language edition}.

Previous research about Wikipedia has tended to focus on the English-language edition. These studies have found that the scientific articles in the English edition compared favorably with Encyclopedia Britannica \cite{giles2005}. However, studies have also suggested the edition suffers from issues of coverage and bias \cite{halavais2008,hecht2009,holloway2007}.

Among these biases, multilingual studies of Wikipedia have revealed each language edition has a self-focus bias \cite{hecht2009}. This bias manifests itself in both the articles users choose to write (and not write) and also in the content of the articles. Hecht and Gergle \citeyear{hecht2010} give an example where the article on psychology in the Spanish-language edition has a section about contributions to the field from Latin America while other  language editions do not.

Even so, a 2011 survey of Wikipedia editors found that just over ``half of Wikipedia editors contribute to more than one language Wikipedia, and an overwhelming majority (72\%) read Wikipedia in more than one language''  (N=4,930).\footnote{\wikisurvey2011}
In addition, Yasseri et al.~\citeyear{yasseri2012-circadian} found registered users from many different timezones contribute to many language editions of Wikipedia. For example, 25\% of edits to the Arabic and Persian editions likely came from users in North America. This suggests diaspora, language learners, or other speakers of these languages play an important role in editing the encyclopedia. Furthermore, the location of these users in North America suggests many of them might speak another language in addition to Arabic or Persian. If so, these users could introduce new information from other language sources and reduce the amount of self-focus in the edition.

Self-focus results were not reported for Arabic or Persian by Hecht and Gergle~\cite{hecht2009}, but the Dutch and Swedish editions were found to exhibit less self-focus. The authors speculated that high bilingualism with English in Dutch and Swedish societies could explain why the Swedish and Dutch editions exhibited less self-focus in their study. They write that users contributing to the Dutch and Swedish editions, ``may  have  gained  significantly  more  guidance from the English Wikipedia, muting their spatial self-focus effect'' \cite[p.~17]{hecht2009}.  This idea, however, is not specifically tested in their paper.

The literature therefore suggests that multilingual users who edit multiple language editions of Wikipedia could play a unique role in diffusing content between different language editions. From seemingly small changes like updated population numbers or new website addresses to large, fast-breaking news developments (\eg the Japanese tsunami and earthquake discussed by Hale \citeyear{hale-chi2012}), multilingual users may help keep content in sync and reduce self-focus bias by introducing new content, updating old content, and correcting errors across multiple language editions.

This \thing{} examines this idea in two ways. First, the articles edited by multilingual users are compared to the articles edited by monolingual users. It is expected that \hypothesis{h:unique}{multilingual users will edit different articles than monolingual users}.
Second, this \thing{} compares the articles edited across language editions by the same user to the network of interlanguage links that link articles on the same concept across language editions. If multilingual Wikipedia users serve as information bridges contributing similar information across multiple editions, then it is expected that \hypothesis{h:bridge}{when a user edits an article in another language that same user will usually also edit the corresponding article in his native language}.

The idea of network effects from network studies or positive externalities from economics may explain in part the reason editors of Wikipedia would contribute to a foreign language edition of the encyclopedia. Network effects suggest that larger-sized platforms or networks have more communicative value than similar, smaller networks. This is obvious in the trivial observation that if only one person in the world had a telephone it would be utterly useless to that person as he would have no one to call. More generally, a social media platform, like a telephone, is only valuable if one's social contacts also use the platform. For without this, a user would have no one with whom to communicate. With each additional social media user, the value of the network grows for the existing users because each person now has a wider array of individuals who they may contact through the network. Crystal~\citeyear{crystal} relates this network effect to languages arguing that the more individuals who use a common language, the more valuable it is for additional individuals to also learn that language. He speculates this effect might account in part for the growth and staying power of English as a global language. This idea is also suggested by Zuckerman \citeyear{zuckerman2013}. 
Similarly, editions of Wikipedia written in more widely spoken languages have the possibility of reaching larger audiences, and past research has  suggested an important factor motivating content production is the extent to which authors believe there is an audience to engage with the content  \cite{zuckerman2008}.

These ideas of network effects related to language size suggest two related hypotheses. The users who cross-language boundaries will \emph{come from} smaller, less-represented languages and will \emph{cross to} larger, more-represented languages.  More specifically, \hypothesis{h:networkFrom}{users writing primarily in smaller-sized language editions will be more likely to cross-language boundaries than users writing primarily in larger-sized language editions}. When these users cross languages, they will most likely cross to a larger-sized language edition (\eg English, German, French). As a consequence of this, \hypothesis{h:networkTo}{larger-sized language editions, English chief among them, will be more likely to have contributions from editors of different languages than smaller-sized language editions}.

\section{Data}

Edits to Wikipedia are broadcast in near real-time over Internet Relay Chat (IRC).\footnote{\url{http://meta.wikimedia.org/wiki/IRC/Channels#Raw_feeds}}
Each edit to any Wikipedia edition is broadcast on the irc.wikimedia.org server on an IRC channel with a name in the format of \#lang.wikipedia (\eg \#en.wikipedia for the English edition of Wikipedia, \#de.wikipedia for the German edition, etc.). Each entry contains the username (or IP address for anonymous users), the title of the article edited, comments written by the user about the edit, the size of the edit (how many bytes larger or smaller the result of the edit is compared to the previous version), and a link to the differences from the previous version. The date and time of the edit is not included, but this information was added by consulting the system clock. Similarly, the IRC channel of the message was recorded to know which language edition the user edited.\footnote{The code used to record the IRC streams (Java), construct the network (Java\slash Hadoop), and perform the analysis (Python/R) are available at \url{http://www.scotthale.net/pubs/?websci2014}.}

All edits for the 46 language editions with 100,000 or more articles%
\footnote{\url{http://meta.wikimedia.org/wiki/List_of_Wikipedias}}
were recorded through IRC from July 8, 2013, to August 9, 2013. Edits to the Simple English edition are excluded for most of the analysis and the role of the Simple English edition is addressed separately in Section \ref{sec:Simple}. In addition to the main, article namespace, Wikipedia has separate namespaces for other content including user pages, portals, and administrative activity. This paper focuses on the main namespace to which the majority of the edits (63\%) were directed. Consistent with prior research, many of these edits (15\% of non-minor edits) were created by bots---automated scripts editing the encyclopedia for consistency, fixing common mistakes, and detecting and reverting vandalism (malicious edits). A number of edits were also from anonymous users without an account (28\% of non-minor edits). Since IP addresses change over time and multiple users may edit from the same IP address, these edits were removed from the dataset. In order to focus on the activity of human editors, only non-minor edits from registered users, who were not listed as bots were considered for further analysis.

Initial analysis of the data suggested that there were many bots operating on the encyclopedia without being officially declared as bots. These suspected bots had very high edit counts across a large number of languages, and human examination of their contributions and user pages suggested most were indeed bots. A number of ideas drawn from the literature were examined and ad hoc subsets of users were manually inspected to arrive at a method to filter these unregistered bots. The most successful approach found was to examine the maximum amount of time between two successive edits from the same user. In accordance with past research, edits for most users (registered bots excluded) were bursty: that is, the edits were clustered such that many edits occurred in small amounts of time separated by comparatively longer absences of edits \cite{geiger2013,yasseri2012-conflicts}. Looking at the length of the longest break between bursts of edits revealed that many users without the bot flag set never had a rest of more than a couple of hours over the entire 32-day data collection window. As most  human editors would need to break longer than this for sleep---and editing activity has previously been shown to follow circadian cycles \cite{yasseri2012-circadian}---these users are likely undeclared bots.

Through manual examination of different thresholds, six hours was chosen. Overall, 114,376 accounts did not have any break in editing of more than six hours over the course of the 32 days in the sample. These users were assumed to be bots (or humans with only one editing burst) and excluded from further analysis. 

One edit is insufficient to determine whether a user edits in multiple languages, while with two edits in two different languages it is unclear which language is the user's primary language. To be certain multilingual users were identified as such and to be able to identify users' primary languages, all users with less than four edits overall (21\%) or less than two edits in their most-edited language (0.6\%) were excluded. As a result, this study focuses on the most active users. This is not, however, a major limitation, as past work has shown the most active users produce a disproportionate majority of the content in the encyclopedia \cite{ortega2008}. 

\subsection{Cross-language alignment}
Previous cross-language studies on Wikipedia relied on the interlanguage links found in each edition of the encyclopedia. These links were maintained by a mixture of humans and machines (bots), but nonetheless contained a number of errors \cite{hecht2009}. The issues were often compounded by having dumps of each language from slightly different dates.

This study uses a new source of inter-language information, WikiData.%
\footnote{\url{http://www.wikidata.org/}}
This new initiative centralizes all interwiki references and category information (and, in the future, statistics and other structured data) in one location. This avoids some previous issues with out-of-date or conflicting interlanguage links. Further study of the impact of the WikiData project on Wikipedia and its editors is not within the scope of this \thing{}, but would be a fruitful area for future research.

When Wikipedia began, each language edition was run independently. User accounts were created separately on each language edition, and thus the same username on different editions may refer to two different persons. As Wikipedia matured, a central authorization system was built to provide for unified login. Unified login allows users to unite their accounts across multiple language editions (and other projects: Wikitionary, Wikiquote, Wikibooks, etc.) and be able to login to all projects and editions at one time. Users who have unified their accounts have ``global accounts'' and information about the user is available from the \emph{Global account manager}.%
\footnote{\url{http://meta.wikimedia.org/w/index.php?title=Special:CentralAuth}}

There was an announcement in April 2013 that any remaining conflicts where different persons had accounts with the same usernames on different editions would be resolved and the accounts renamed.%
\footnote{\url{http://meta.wikimedia.org/wiki/Single_User_Login_finalisation_announcement}}
This was to take place in May 2013, but was delayed first to August 2013 and then to an unspecified future date. Once this step is taken it will be trivially easy to determine if one user edits multiple language editions. At the point of data collection, however, it still remained technically possible for one person to have two differently named accounts on different editions, or for two persons to have accounts of the same name on different editions.

The publicly-available data makes it difficult to identify one person with multiple accounts (false-negative monolingual). It is possible, however, to check whether a given account is a global account. If it is a global account, it is possible to get a list of all the language editions on which the user is active. This makes it possible to avoid any false-positive multilingual user classifications.

For this study, all usernames were first assumed to be unique across the editions. The usernames editing multiple language editions were identified and classified as possible multilingual users. Each of these usernames was checked against the \emph{Global account manager} to ensure that the user was a global user registered with all the language editions the user was recorded as having edited during the data collection period. Users who were not registered as global users or whose global username was not associated with all the editions the user was recorded as having edited were treated as separate users.

There were very few false-positive matches found. Only 572 usernames found to be editing multiple editions were not global accounts. In a further 50 cases, a global username existed but was not associated with all language editions where the username was used. These local, non-global users were treated as separate users. The small number of these matches means that this correction has minimal  effect on the results.

The available data does not easily allow the discovery of false-negatives---one user with different usernames on different editions. In addition, it is not possible to know if a user reads multiple language editions, while editing only one edition. Therefore, the number of multilingual users presented in this \thing{} is a lower bound on the actual amount of multilingual activity happening on Wikipedia.

\section{Analysis}
\begin{table}[tb]
\begin{center}
\begin{tabular}{lrrrrr}
  \toprule
  \multicolumn{1}{p{0.15\columnwidth}}{Language} &
  \multicolumn{1}{p{0.1\columnwidth}}{Edits} &
  \multicolumn{1}{p{0.1\columnwidth}}{Articles} &
  \multicolumn{1}{p{0.1\columnwidth}}{Users} &
  \multicolumn{1}{p{0.1\columnwidth}}{NP users} &
  \multicolumn{1}{p{0.1\columnwidth}}{NP edits} \\
  \midrule

        English & 1,389,647 &   518,405 &    27,476 &        18\% &         3\% \\
         German &   256,495 &   125,647 &     5,967 &        18\% &         2\% \\
         French &   250,828 &   106,027 &     4,549 &        25\% &         3\% \\
         Spanish &   191,934 &    66,848 &     4,338 &        24\% &         3\% \\
         Russian &   239,267 &    92,326 &     3,961 &        16\% &         1\% \\
         Japanese &   106,848 &    56,406 &     3,551 &        11\% &         2\% \\
         Italian &   160,191 &    69,534 &     2,919 &        25\% &         2\% \\
         Chinese &   112,888 &    42,937 &     2,309 &        14\% &         1\% \\
         Portuguese &    67,505 &    32,753 &     1,730 &        29\% &         4\% \\
         Dutch &    80,535 &    39,463 &     1,500 &        33\% &         3\% \\
         Polish &    67,038 &    37,393 &     1,454 &        30\% &         3\% \\
         Swedish &    42,390 &    25,269 &       904 &        43\% &         4\% \\
         Ukrainian &    54,241 &    22,537 &       898 &        36\% &         3\% \\
         Hebrew &    37,889 &    13,224 &       832 &        16\% &         2\% \\
         Arabic &    43,924 &    15,993 &       729 &        20\% &         3\% \\
   \bottomrule
\end{tabular}

\caption{Statistics for the top 15 language editions in the sample. The \emph{Users} column includes all users who edited the edition during the data collection period. A  percentage of these users (\emph{NP users}) are non-primary users who edited a different language edition more frequently. \emph{Edits} and and \emph{NP Edits} are defined similarly.}
\label{tbl:summary}

\end{center}
\end{table}

\begin{figure}
	\begin{center}
		\includegraphics[width=.4\textwidth]{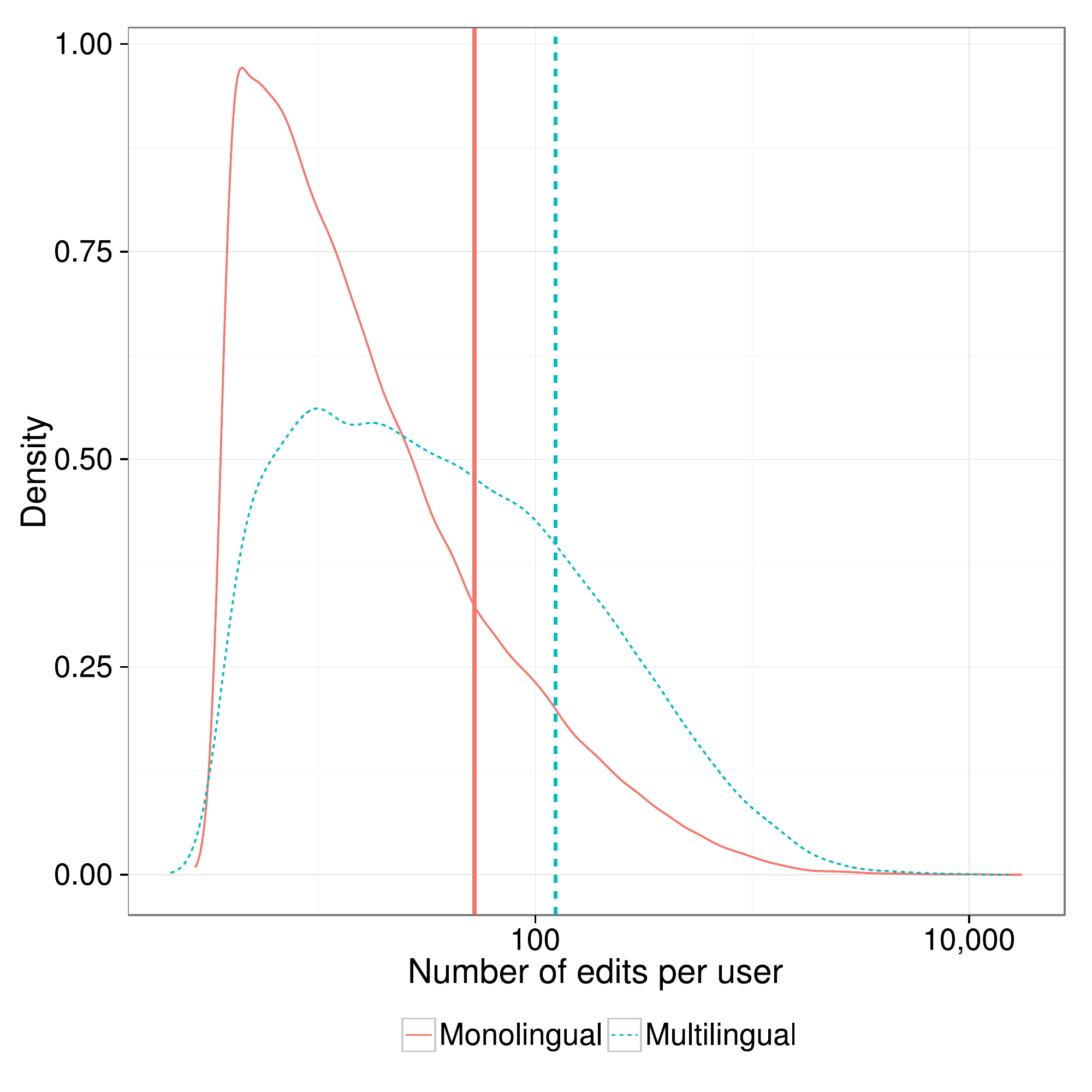}%
		\caption{Density plot comparing the number of edits made by monolingual and multilingual Wikipedia users.}
		\label{fig:wikimultimono}
	\end{center}
\end{figure}

Excluding the Simple English edition, 55,568 registered, human users edited at least one edition of Wikipedia two times or more and had at least four edits across all editions during the 32-day data collection period. This resulted in a total of 3,518,955 edits. Most of these edits (39\%) were to the English language edition. Similarly, most users (40\%) edited the English-language edition of the encyclopedia more than any other edition (Table \ref{tbl:summary}).

Consistent with H\ref{h:wikihomophily} a relatively small number of users (8,544 or 15.4\%) edited multiple editions of the encyclopedia. These users were categorized as \emph{multilingual} while all remaining users were classified as \emph{monolingual}. Multilingual users were significantly more active than their monolingual counterparts. Multilingual users made a mean 124 (median 32, sd 299) edits overall, while monolingual users made only a mean 52 (median 13, sd 192) edits overall (Figure \ref{fig:wikimultimono}). These additional edits by multilinguals are not only in other language editions but also in each user's primary language edition that the user edited most frequently. Multilinguals made a mean 113 (median 26, sd 285) edits to their primary language editions of the encyclopedia. Indeed, while only 15.4\% of all users, multilingual users were responsible for 30.1\% of all edits captured during the month.

Multilingual users were not just editing the same articles more, but also edited a wider number of articles. Multilingual users edited a mean 69 (median 16, sd 191) articles while monolinguals edited a mean 27 (median 5, sd 133) articles. Logically following from the fact that multilingual users were more active in their primary languages than monolingual users, it is clear multilinguals were not more active simply because they had more articles across more languages they could have edited. As discussed in the next subsection, multilinguals only directed a small percentage of their edits to their non-primary languages. Multilingual users were still more active than their monolingual counterparts after collapsing together articles in different languages on the same concept as determined by interlanguage article links. In this way, for example, editing United States (English) and Estados Unidos (Spanish) only counted as editing one ``concept'' since the two articles are linked together by interlanguage article links. Monolingual users edited the same number of concepts as articles since they only edited one language, while multilingual users edited a mean 65 (median 15, sd 185) concepts. All of these differences are significant as established with two-tailed t-tests ($p < 2.2 \times 10^{-16}$).

In contrast, the size of the edits made by multilinguals and monolinguals do not differ significantly. Edits by multilinguals had a mean size of 331 bytes (median 143, sd 912), while edits by monolinguals had a mean size of 339 bytes (median 125, sd 1327).

\subsection{What do multilinguals edit?}
\begin{figure}
	\begin{center}
		\includegraphics[width=.4\textwidth]{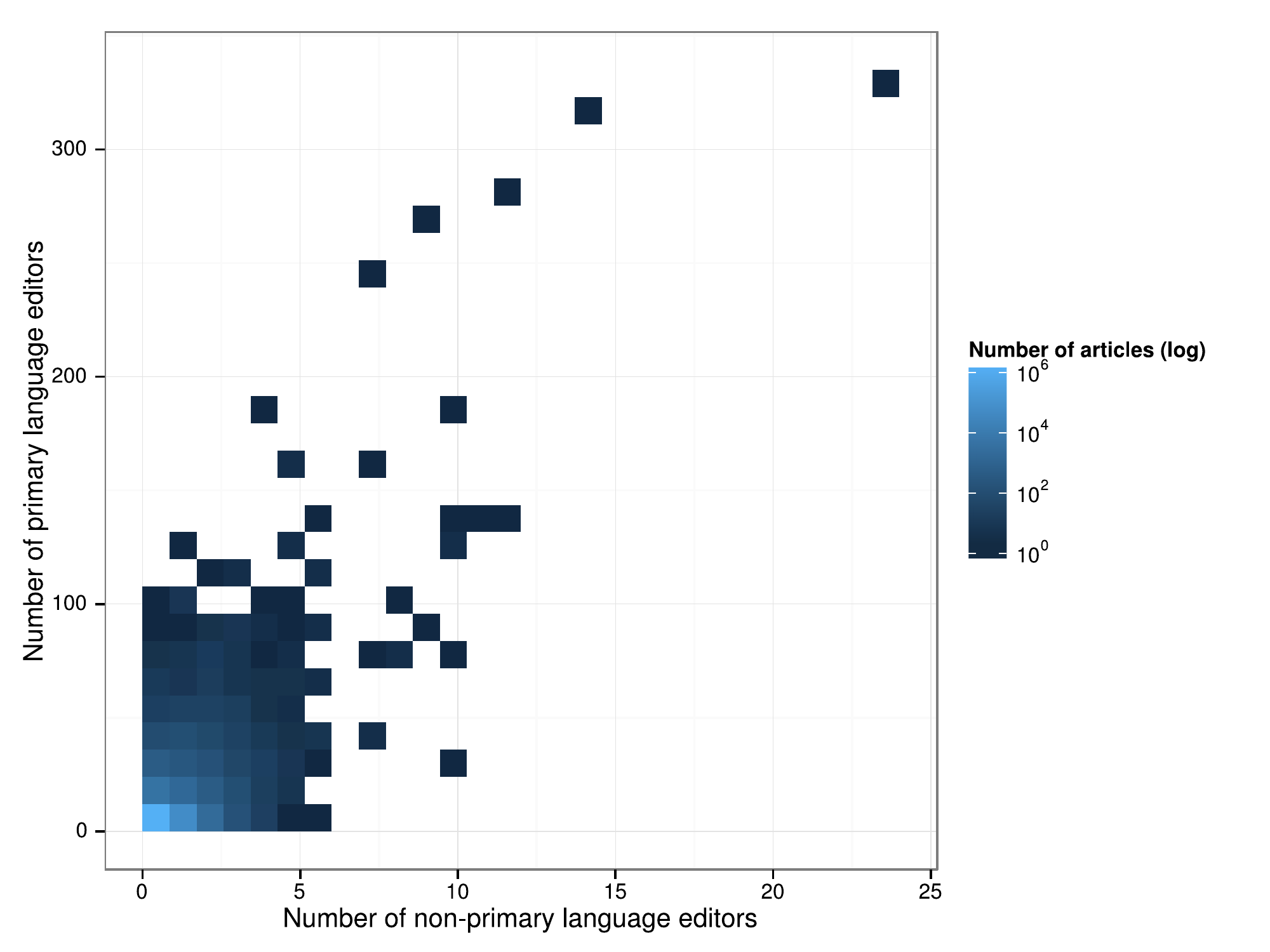}
		\caption{2D density plot of the number of multilingual users editing articles in a non-primary language against the number of monolingual users editing the articles.}
		\label{fig:homeForeignArticleDistribution}
	\end{center}
\end{figure}

Given the low overlap in content between language editions of Wikipedia, multilingual users may offer unique contributions to the editions they edit. This section examines the edits of multilingual users to their non-primary language editions (that is, to editions other than the editions they edited most frequently).

Edits from multilingual users writing in their non-primary language are an extremely small fraction of all edits to Wikipedia. Only 2.6\% of edits are from users writing in their non-primary languages. To some extent, multilingual users edit similar articles in their non-primary languages as do monolingual users. The 2D density plot in Figure \ref{fig:homeForeignArticleDistribution} shows that the articles with the largest number of non-primary users also have a large number of primary users. There's a positive correlation of 0.25 between the number of multilingual users editing an article in a second language and the number of monolingual users editing the article. The most dense region is near the origin where most articles are edited by a small number of users. Within this region, however, multilingual users are often editing articles not edited by other users: 44\% of the articles edited by multilingual users in their non-primary languages were not edited by any monolingual user during the data collection period.

\begin{figure}
	\begin{center}
		\includegraphics[width=.4\textwidth]{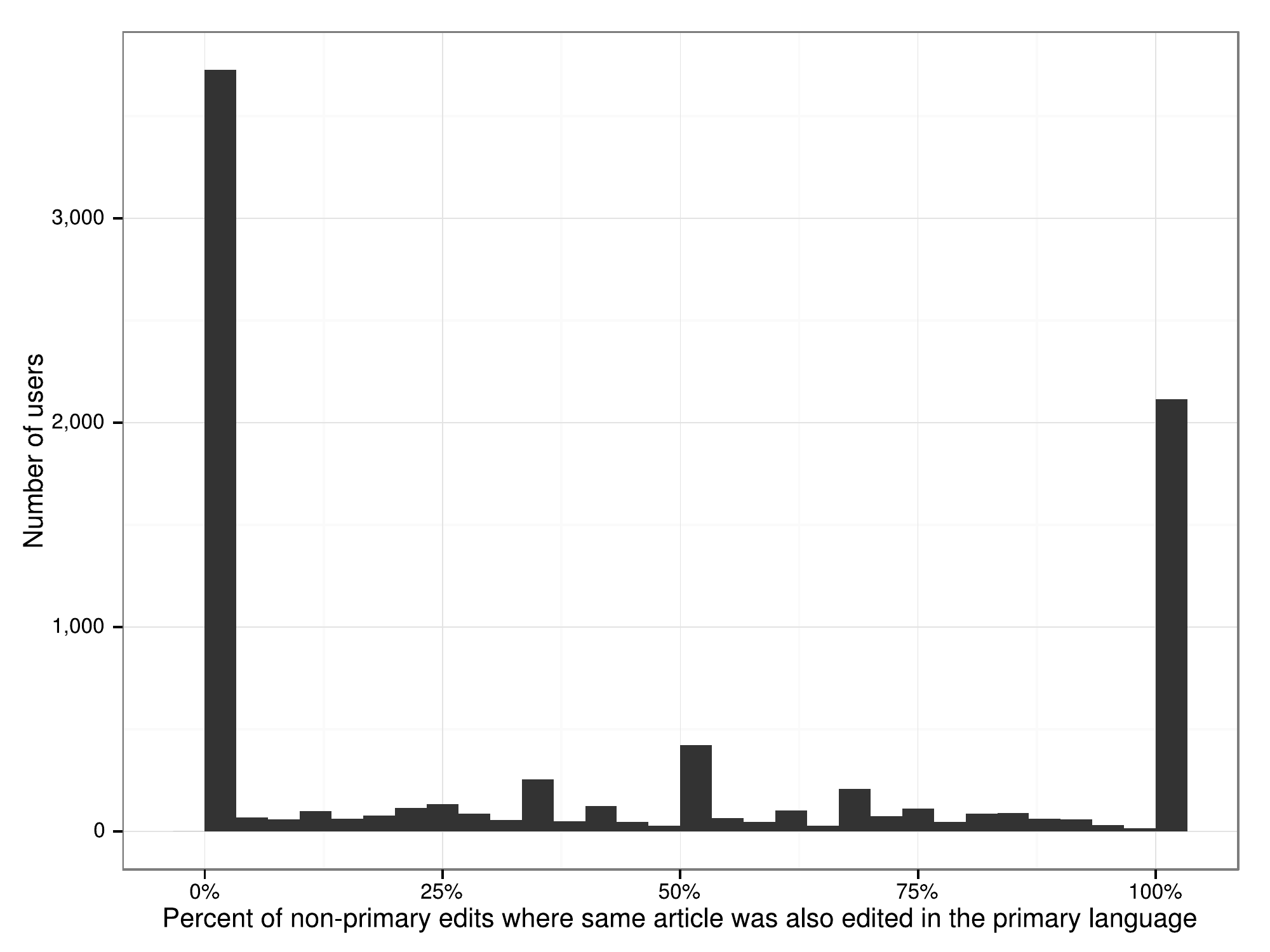}%
		\caption{Histogram showing the distribution with which multilingual users edited articles in other languages that they also edited in their primary languages. The distribution is bimodal. A large number of users did not edit any of the same articles in their primary languages, but a large number of users always edited the same articles in their primary languages.}
		\label{fig:homeForeignUserDistribution}
	\end{center}
\end{figure}

Using the WikiData information for interlanguage article links, it is possible to connect articles across languages (\eg the English-language \emph{United Nations} article is on the same concept as the Spanish-language \emph{Organizaci\'on de las Naciones Unidas} article). This makes it possible to check when a multilingual user edited an article in a non-primary language, if that user also had edited the equivalent article in his or her primary language. Overall, 44.5\% of the edits to non-primary languages by multilingual users were to articles where the user had edited the same article in his or her primary language. The underlying distribution per user, however, is bimodal at the two extremes (Figure \ref{fig:homeForeignUserDistribution}). 43\% of multilingual users did not edit the equivalent articles in their primary languages at all. On the other hand, 25\% of multilingual users always edited the equivalent articles in their primary languages.

Part of this behavior is explained by the fact that some of the articles edited by multilingual users in their non-primary languages did not exist in their primary languages. Overall, 73\% of the articles edited by multilingual users in a non-primary language existed in the primary languages of those users. Ignoring the instances where an equivalent article did not exist in the users' primary languages, 59.8\% of edits to non-primary languages by multilingual users were to articles the user had edited in his or her primary language. The distribution remains bimodal with 34\% of users not editing the equivalent articles in their primary languages, and 37\% of users always editing the equivalent articles in their primary languages.

While the size of edits made by multilingual and monolingual users did not differ significantly, the size of edits multilingual users made in their non-primary languages were significantly smaller than the edits they made in their primary languages. Considering only edits with a positive size (\ie not edits that removed more text than was added) multilingual users made edits with a mean size of 569 bytes (median 260, sd 1327) in their primary languages and a mean size of 468 bytes (median 83, sd 2156) in their non-primary languages. Nonetheless, 25\% of multilingual users actually made larger positive-sized edits in their non-primary languages as compared to positive-sized edits in their primary languages.

Comparisons of edit sizes across languages is difficult for two reasons: first, different characters require a different number of bytes to store, and second, the information content contained in one character differs across languages. One standard English character is usually one byte, while a special or accented character (\eg {\'a}) is usually two bytes, and a character from a more complex language like Japanese, Chinese, or Korean is generally three bytes. In contrast, however, one English character usually contains less information content than one Japanese or Chinese character, which could represent a full word. An information-theoretic approach, using entropy, has previously been employed to compare the information content per character across different languages on Twitter \cite{neubig2013}, and a similar approach could be employed to compare Wikipedia edit sizes across languages. Such an approach, however, would require the content of the edits rather than the meta-data about edits used here.

Overall, these findings support H\ref{h:unique} that multilingual users would make unique contributions to the encyclopedia by editing articles less edited by monolingual users. The data is mixed for H\ref{h:bridge} which suggested multilingual users would often edit the same article in their primary and non-primary languages. For a quarter of users this was always true.  However, just over two-fifths of multilingual users did not edit the equivalent articles in their primary languages. Data on the articles users view is not available to know whether these users viewed the equivalent articles in their primary languages before editing in another language.

\subsection{Variations by language}
\begin{figure}
	\begin{center}
		\includegraphics[width=0.5\textwidth]{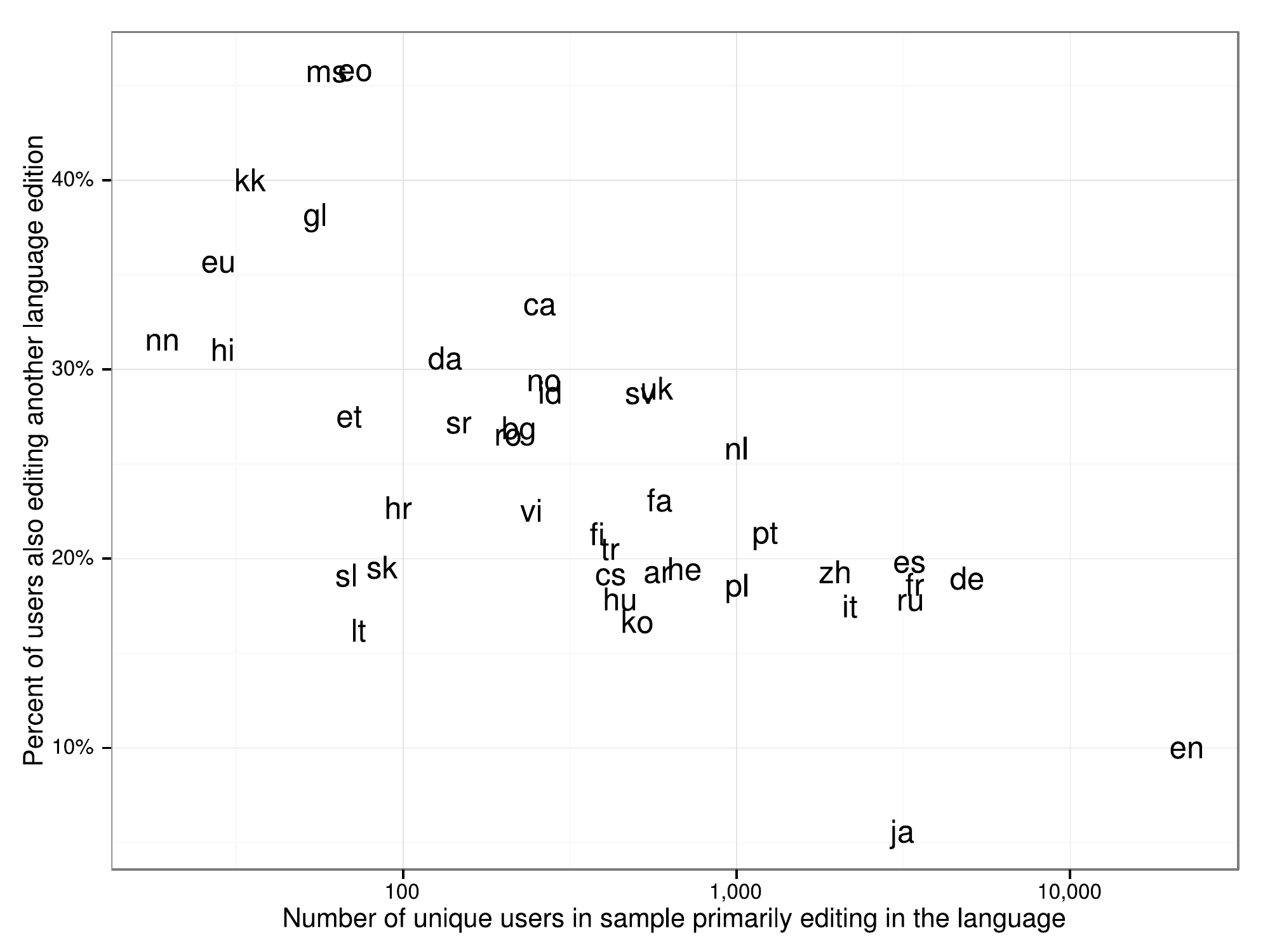}
		\caption{Scatter plot of language size (number of unique users) and percentage of users who are multilingual (edit more than one language edition). The three editions with less than 10 users in the sample are omitted (Uzbek, Cebuano, and Waray-Waray).}
		\label{fig:introversion}
	\end{center}

\end{figure}

The percentage of users classified as multilingual varied across the language editions studied. Previous research suggested this variation would correlate with the total number of users and/or the number of articles in each edition. Figure \ref{fig:introversion} shows the percentage of users primarily editing each language edition that also edited a second language edition compared with the total number of users primarily editing the language edition. Consistent with the suggestions of prior research, there is a strong correlation between the two variables. Looking only at languages with at least 10 users in the sample to avoid small number issues, the log of the number of users primarily editing each edition and the percentage of users editing multiple editions are correlated with a coefficient of $-0.69$. Similar results hold for comparing the percentage of users who are multilingual to the number of articles in each language edition, where the correlation coefficient is $-0.46$. (These two measures are interdependent as the total number of articles per edition and the number of users in the sample per edition have a correlation coefficient of 0.90.)

Among the smallest-sized editions, Esperanto (eo) and Malay (ms) stand out as two languages with high levels of multilingualism among their primary editors. It is surprising that Esperanto was not higher given that it is a constructed language and thus has no native speakers. Nonetheless, nearly 46\% of the editors of the Esperanto edition edited that edition more than any other. Italian (it), Slovenian (sl), and Slovak (sk) are similarly sized but with far lower levels of multilingualism.

Among larger-sized languages, Catalan (ca), Swedish (sv), Ukrainian (uk), and Dutch (nl) all had relatively high levels of multilingualism. In contrast, the lowest level of multilingualism is found among users primarily editing the Japanese (ja) language edition, where only 6\% of the users edited another edition.

While some exceptions emerge, the findings support H\ref{h:networkFrom}: in general a larger percentage of the users primarily editing smaller-sized editions are multilingual.

\subsection{Language crossings}

Hypothesis H\ref{h:networkTo} predicted that when users did edit a second edition, that edition would almost always be English or, to a lesser degree, another large edition. In order by the number of users active in the last 30 days, the largest editions of Wikipedia are English (129,900 active users), German (20,300), Spanish (15,800), French (15,500), Japanese (11,400), Russian (10,800).%
\footnote{The latest information on the number of users and articles for all editions of Wikipedia is online at \url{http://meta.wikimedia.org/wiki/List_of_Wikipedias}}

\begin{figure}
	\begin{center}
		\begin{subfigure}{\columnwidth}
			\includegraphics[width=\textwidth]{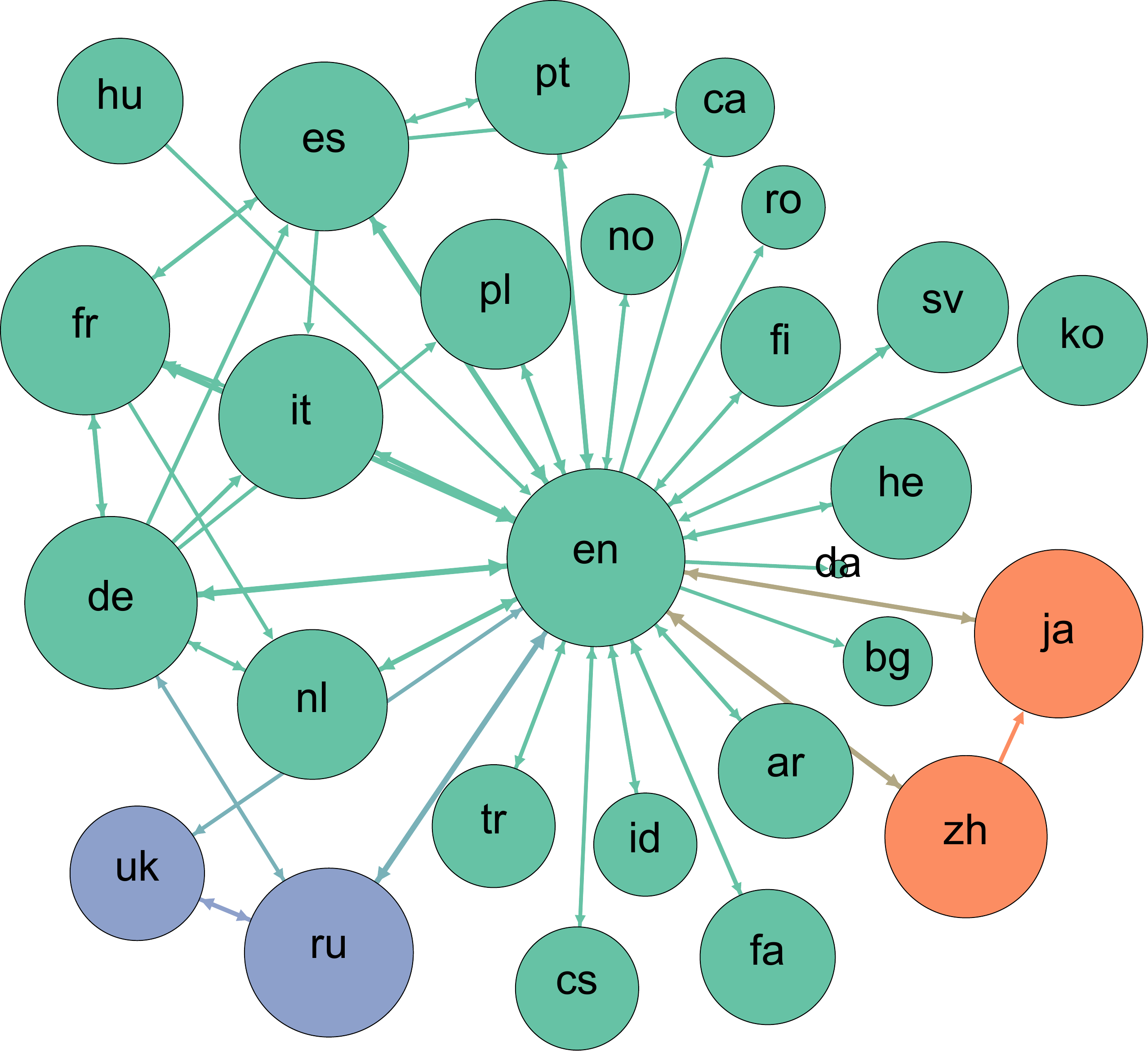}
			\caption{Network graph with English}
			\label{fig:networkEn}
		\end{subfigure}
		\begin{subfigure}{\columnwidth}
			\includegraphics[width=\textwidth]{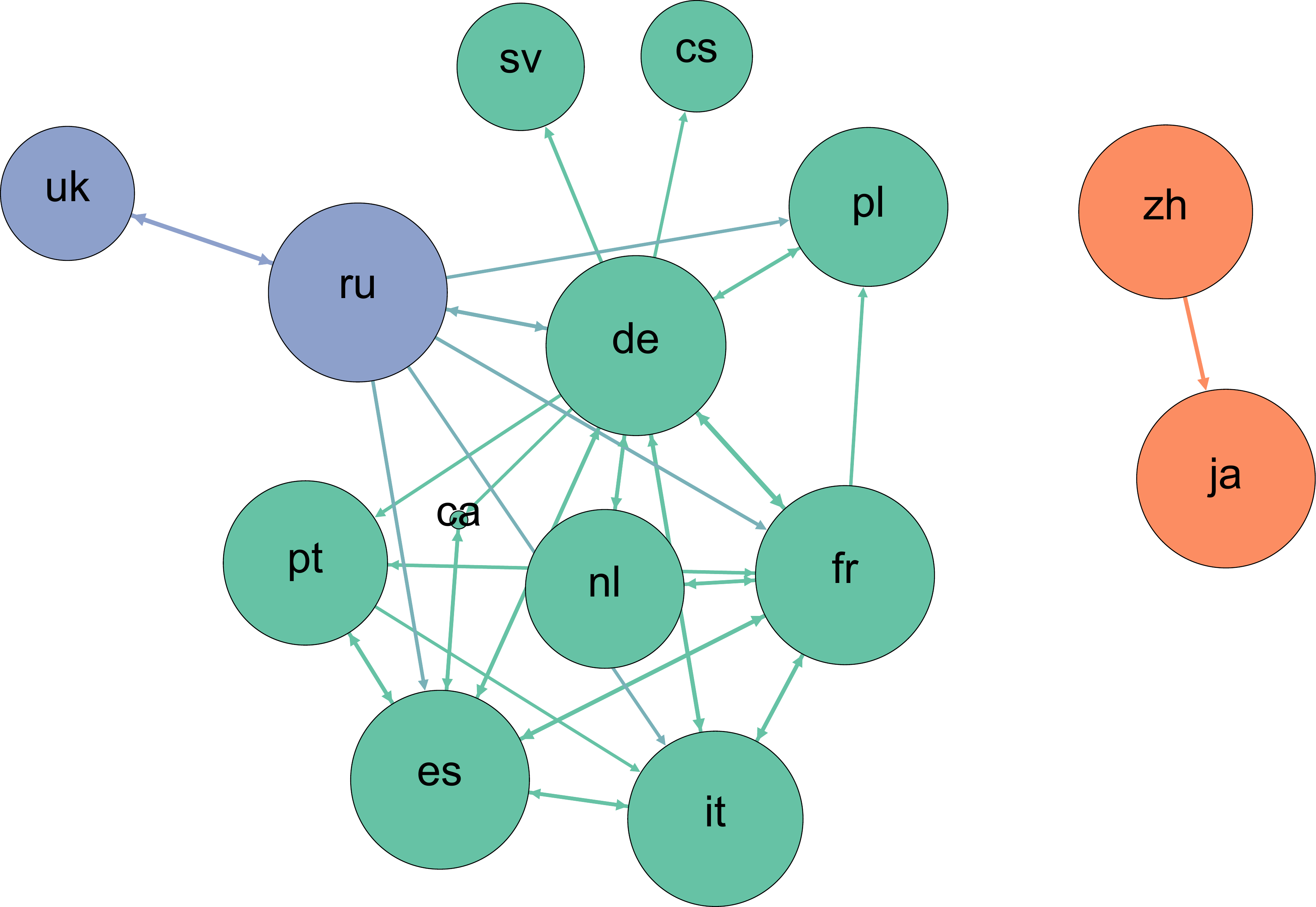}
			\caption{Network graph with English removed}
			\label{fig:networkNoEn}
		\end{subfigure}
	\caption{Network graphs of co-editing patterns. Nodes represent language editions of the encyclopedia and the directed, weighted edges show the log of the number of users primarily editing one language edition who edited another edition as well. Both graphs show only edges with weights over 1.96 standard deviations above the mean. The top graph shows all language editions. In the bottom graph, the English edition is removed and the distribution of edges recalculated. Colors indicate communities found by the infomap community detection algorithm.}
	\label{fig:network}
	\end{center}
\end{figure}

The bipartite network of users and articles was collapsed to a network of language relationships. Each article was assigned to the node corresponding to the language edition to which it belonged. Similarly, users were grouped with the node representing the language they edited most frequently. Each directed, weighted edge $e_{ij}$ records the log of the number of editors primarily editing language edition $i$ that also edited language edition $j$. This network is shown in Figure \ref{fig:networkEn}.

The number of users represented by each edge ranged from 1 to 775 with a mean value of 15.3 and a standard deviation of 50.7. Only edges with a log value at least 1.96 standard deviations above the mean of all log values are shown on the graph. This corresponds to edges with 60 or more users. Isolates (languages unconnected to any other language) are removed from the network diagram. Note that in contrast to the previous section, the network graph shows the logarithm of the number of users editing multiple editions and not the percentages of users editing each edition that also edit another edition.

The network reveals the English edition (en) does receive a large amount of attention from multilingual users in other languages. Every node in the graph is connected to English. Most of these edges are reciprocal, but in three-quarters of the cases more users from another language edited English than users from English edited the other language. Despite the very large  size of English, this also holds globally as 4,659 users from other languages edited English while only a total of 3,673 primary users of English edited another language. When users primarily editing the English edition did edit another language edition, the largest number of users edited the Spanish (es), German (de), and French (fr) editions.

There are only four languages that have a directed edge from English that is not reciprocated. These languages are Romanian (ro), Danish (da), Bulgarian (bg), and Catalan (ca). Each of these four languages is quite small, and while a sizable percentage of the users primarily editing these language editions also edited English%
\footnote{27--33\% of the users in each of these four languages edited multiple editions. Most of these multilingual users edited English in the case of Romanian (95\%), Danish (88\%), and Bulgarian (88\%). Of the multilingual users that primarily edited Catalan, 50\% also edited English.}
they simply did not constitute sufficient volume to rise above the edge weight threshold and appear on the graph.

There are are some strongly connected language pairs not involving English. German users edit the French edition, and Russian and Ukrainian users edit each others' editions with some frequency. Figure \ref{fig:networkNoEn} shows the same network, but with English removed. The edge weight distribution is recalculated and edges with 33 or more users are shown (corresponding to 1.96 standard deviations above the mean of the log values with English removed).

Even with the English edition removed, editions with a larger number of active users continue to structure the network. The second-largest edition, German, is connected to every node except Ukrainian (uk), Japanese (ja), and Chinese (zh).

The infomap community detection algorithm \cite{rosvall2009,rosvall2008} finds the same community structure with and without English as shown with node color in Figure \ref{fig:network}. The largest community is centered around the largest  language, English or German. A strong relationship is present between Ukrainian and Russian (ru) where Ukrainian users edit Russian and English but rarely another language edition. Similarly, Chinese users edit Japanese and English but rarely any other edition. Unlike the Russian\slash Ukrainian relationship, the edge from Chinese to Japanese is one way. Indeed, apart from editing English, Japanese users rarely edit any other language.

With English removed, it is also worth noting that all the romance languages (Spanish [es], Italian [it], French [fr], and Portuguese [pt]) have mutual edges between them. The only exception is Catalan (ca), which is only connected to Spanish and German. Nonetheless, the many links to German from other language editions overshadow these connections in the community detection algorithm.

These findings support hypothesis H\ref{h:networkTo} that multilingual users from smaller languages would mostly cross language boundaries to edit larger-sized languages. English receives edits from users in almost every other edition. Even with English removed, the second-largest edition, German, receives edits from users in a large number of other editions. Nonetheless, regional and linguistic patterns are also evident in the co-editing network.

\subsection{The role of Simple English}
\label{sec:Simple}

The Simple English edition of Wikipedia (hereafter Simple) is written in English, but aims to use simpler grammar and shorter sentences. While intended to be primarily read by children, adults with learning difficulties, and second-language learners of English, past research has indicated it may also be edited by many speakers of other languages \cite{yasseri2012-circadian}.

The findings presented so far in this \thing{} have excluded all edits to Simple. Its inclusion, however, makes little difference to the findings. There is a strong link between Simple and the English edition. Of the 221 users in the data sample who edited Simple, over half (124 users) primarily edited English. Users editing Simple and primarily editing a non-English edition were spread thinly over 26 other languages. The four largest of these languages were German (13 users), French (10), Dutch (9), and Russian (8). Two-thirds of the users editing Simple were multilingual users, who already edited at least two other language editions. Of the 76 users editing Simple who were previously classified as editing only one edition,  66 were primarily editors of the English edition. Simple makes little difference to the structure of the co-editing networks in Figure \ref{fig:network}. Enough English users edit Simple to have a small directed edge from English to Simple, but no other edges are added.

Like the Esperanto edition, there appears to be a cohort of users dedicated more to Simple than to their native languages. There are 21 users in the dataset who edited Simple more than any other edition (14 of these users edited the English edition second-most). In addition, there are a further 44 users not in the dataset, who edited Simple more than two times but did not have at least four edits across all editions when their edits to Simple were excluded. Just under half (45\%) of all the users who edited Simple most often did not edit any other edition at all. Of those users that did edit a second edition less frequently than Simple, that edition was English for all but 9 users.

There have been two proposals in the past to close Simple, both of which have failed. Whatever the utility of the edition to readers, it does have a dedicated community of editors. In this respect, it is very similar to the Esperanto edition, where 54\% of the users that primarily edited Esperanto edited no other language edition.

\section{Discussion}

By far, most Wikipedia users edited only one language edition, confirming H\ref{h:wikihomophily}. However, just over 15\% of users also edited multiple language editions. These multilingual users were found to be more active than their monolingual counterparts making more than 2.3 times as many edits per user on average. Most of this additional activity occurred in the users' primary languages, with only 2.6\% of all edits being made by users in their non-primary languages. It is important to note that this is a correlation between multilingualism and activity and not causation. It may be that the most dedicated and active users of Wikipedia contribute to multiple language editions regardless of how great their foreign-language skills really are.
Survey work, for instance, has shown many Internet users in Uzbekistan engaged with foreign-language content even while simultaneously reporting low comfort with foreign languages \cite{wei}.
Regardless of the direction of this relationship, it will be important to keep these multilingual users in mind when considering design changes to Wikipedia.

The percentage of users editing multiple languages on Wikipedia is similar to the 11\% of users found to tweet in multiple languages on Twitter \cite{hale-chi2014}. On the other hand, the percentage of users editing multiple language editions is far less than the 50\% of users that self-reported editing multiple language editions in the 2011 Wikipedia editors survey.\footnote{\wikisurvey2011} This could perhaps follow from the idea that the most dedicated users are multilingual and thus more likely to take the time to respond to a survey about Wikipedia when given the opportunity. Alternatively, given that multilinguals only made a small fraction of their edits to their non-primary language editions, it is possible that more users would be observed editing multiple language editions if they were observed for a longer period of time.

Multilingual users editing more than one edition of Wikipedia can bring information, sources, and perspectives from the primary edition they edit to other editions. A large portion (44\%) of the articles edited by multilinguals in their non-primary languages were to articles that no monolingual users in that language edited during the month of study. A similar percentage of all edits by multilinguals in their non-primary languages were to articles that the same multilingual user had edited in his or her primary language. This suggests that multilingual users are making unique contributions not duplicated by monolingual users and that in many cases multilingual users are working on the same article in multiple languages.

Hecht and Gergle \citeyear{hecht2009} have previously suggested that users crossing between different languages like this might reduce the amount of self-focus bias in Wikipedia. They found the Dutch and Swedish editions to be less self-focused than other editions. The research presented in this \thing{} supports their conjecture that this is likely due to higher levels of multilingualism among speakers of these languages. This research shows that a relatively higher percentage of users primarily editing the Dutch or Swedish editions also edit another language edition. Hecht and Gergle \citeyear{hecht2009} also found the Portuguese edition to be less self-focused. The rate of multilingualism among users primarily editing the Portuguese edition in this study is slightly above the mean, but is mostly explained by the size of the Portuguese edition. Overall self-focus bias among the 15 editions studied by Hecht and Gergle \citeyear{hecht2009} is negatively correlated with the level of multilingualism found in this study. That is higher levels of multilingualism are generally associated with less self-focus bias. The correlation coefficient between the measures is $-0.67$, although this drops to $-0.33$ if English and Japanese are excluded. Multilingualism is one of perhaps several factors affecting the level of self-focus bias in different editions of Wikipedia, and this study has only been able to observe cross-language editing and not cross-language reading. Further study should identify additional factors affecting self-focus bias and their relative roles.

Multilingual users are found in all language editions. Generally, however, a higher percentage of users primarily editing smaller-sized language editions are multilingual compared to users primarily editing larger-sized editions, supporting H\ref{h:networkFrom}. This is also consistent with prior qualitative and survey work. Of the outliers found, Esperanto and Malay had higher percentages of multilingual users than their sizes would predict, while Japanese had a much lower percentage than its size would predict. Malay users have previously been found to be among the most multilingual user groups on Twitter, while Japanese users were similarly found to be the least multilingual group on Twitter \cite{hale-chi2014}. This points to the importance of language-specific factors, which are also shown in the rather simple case of Esperanto being a constructed language with no native speakers.

Differences between the results on Twitter \cite{hale-chi2014} and those found here on Wikipedia also suggest platform-specific characteristics affect the levels of multilingualism users exhibit. For example, in the Twitter study Italian users were more multilingual on Twitter than their size suggested, while the opposite was found here.
In addition, the correlation in the Twitter study between language size and levels of multilingualism was very weak whereas the correlation on Wikipedia was much stronger. Further research will be needed to untangle the role of design and platform-specific characteristics affecting the levels of multilingualism on different platforms.

When users did edit a second language edition, that edition was most often English, supporting H\ref{h:networkTo}. English users did edit many other language editions, but these users were a much small percentage of the English user total than the percentage of users primarily editing other languages that also edited English. Even with English removed, the network of language crossings was centered around German, the second-largest edition. Some regional and linguistic groups were also apparent, pointing towards the importance of geo-linguistic factors \cite{liao2010} in the cross-language activity of Wikipedia users.

Including or excluding the Simple English edition of Wikipedia had little impact on the findings of this \thing{}. Many users editing Simple already edited two other editions and were classified as multilingual. Past research analyzing the location of users through the circadian rhythms of their edits found more editors of Simple were in Europe and the Far East\slash Australia compared to the English edition \cite{yasseri2012-circadian}. This raised speculation that English as a second language (ESL) speakers might edit Simple more than the English edition. This research finds, however, that the English edition is edited much more by users of other languages than is Simple. Thus, the difference in geographic spread between the two editions is more likely one of awareness and commitment to editing among English speakers rather than a native\slash ESL divide. Indeed, this research has shown that like Esperanto, there is a dedicated editing community for Simple. Many users edit Simple (or Esperanto) more than any other edition despite no one being a native speaker of Simple (or Esperanto).

Overall, this study shows multilingual users play a unique role on Wikipedia editing articles different to those edited by monolingual users. Multilingual users may further transfer information between language editions and thereby reduce the levels of self-focus bias in the encyclopedia. The correlation between self-focus bias and multilingualism is present, but noisy, and further research is needed to identify other factors that also affect self-focus bias. Finally, differences between the levels of multilingualism by language previously found on Twitter with the levels found in this research on Wikipedia indicate design and platform-specific factors that future research should explore.

\section{Acknowledgments}
I would like to thank Taha Yasseri, Eric T. Meyer, Jonathan Bright, and Mike Thelwall,
as well as the anonymous reviewers
who provided helpful comments on previous versions of this \thing{}.

\bibliographystyle{abbrv}

\end{document}